\documentclass[fleqn,usenatbib]{mnras}

\usepackage{newtxtext,newtxmath}

\usepackage[T1]{fontenc}

\DeclareRobustCommand{\VAN}[3]{#2}
\let\VANthebibliography\thebibliography
\def\thebibliography{\DeclareRobustCommand{\VAN}[3]{##3}\VANthebibliography}

\usepackage{graphicx}	
\usepackage{amsmath}	
\usepackage{booktabs}

\newcommand\grp{\ensuremath{G_\mathrm{RP}}}

\newcommand\g{\ensuremath{G}}
\newcommand\gaia{\textit{Gaia}}

\newcommand\teff{$T_{\mathrm{eff}}$}

\title[Resolving J1250+0455AB a UCD binary] { J1250+0455AB an ultracool binary in a hierarchical triple system}

\author[Sayan Baig]{
Sayan Baig$^{1,2}$\thanks{E-mail: s.baig@herts.ac.uk},
R. L. Smart$^{2,1}$,
Hugh R.A. Jones$^{1}$,
E. Pinna$^{3,4}$,
A. Sozzetti$^{2}$,  
Gemma Cheng$^{1}$,
\newauthor \phantom{} 
Felice Cusano$^{3,4}$,
Fabio Rossi$^{3,4}$,
Cedric Plantet$^{3,4}$ and
Guido Agapito$^{3,4}$
\\
\\
$^{1}$ School of Physics, Astronomy and Mathematics, University of Hertfordshire, College Lane, Hatfield, AL10 9AB, UK\\
$^{2}$ Istituto Nazionale di Astrofisica, Osservatorio Astrofisico di Torino, Strada Osservatorio 20, I-10025 Pino Torinese, Italy\\
$^{3}$ Istituto Nazionale di Astrofisica, - Osservatorio Astrofisico di Arcetri, largo E. Fermi 5, 50125 Firenze, Italy \\
$^{4}$ ADONI - ADaptive Optics National lab in Italy, Italy}

\date{Accepted XXX. Received YYY; in original form ZZZ}

\pubyear{2024}

\begin{document}
\label{firstpage}
\pagerange{\pageref{firstpage}--\pageref{lastpage}}
\maketitle

\begin{abstract}
 We report the discovery of the ultracool dwarf binary system J1250+0455AB, a low-mass (M$_\odot$$_\mathrm{tot} <$ 0.2 M$_\odot$) system in which the components straddle the M/L dwarf boundary. The binary was resolved through near-infrared adaptive optics imaging with LUCI1-SOUL on the Large Binocular Telescope, revealing a projected angular separation of 0.17 $\pm$ 0.015$\arcsec$, which, combined with a system distance of  $71 \pm 5.8$\,pc, corresponds to a physical separation of 12.2 $\pm$ 1.5\,AU at a position angle of 84.8 $\pm$ 0.2°. We estimated the orbital period of J1250+0455AB to be 156 $\pm$ 8\,yr,  the bolometric luminosities of the primary and secondary luminosities as $\log (L_\mathrm{bol} / L_\odot) = -3.45 \pm 0.04$ and $-3.58 \pm 0.04$, respectively, with the spectral types of M9 and L0 determined through binary template fitting and spectrophotometric relations. This binary system is part of a hierarchical triple with a separation of 10.44$\arcsec$ from its primary. We estimated the age of the system from the rotational period of the primary star as $0.56^{+0.07}_{-0.06}$ Gyr. Using evolutionary models, for each component we estimate the mass [0.079 $\pm$ 0.002\,M$_\odot$ / 0.072 $\pm$ 0.003\,M$_\odot$], effective temperature [2350 $\pm$ 38\,K / 2200 $\pm$ 43\,K], and radius [0.113 $\pm$ 0.003\,R$_\odot$ / 0.108 $\pm$ 0.002\,R$_\odot$]. Based on the system's binding energy, total mass, and separation, J1250+0455AB is predicted to be a highly stable system, remaining bound for $>$ 10\,Gyr. J1250+0455AB extends the growing population of UCD benchmark systems, providing a new system for refining evolutionary theories at the lowest stellar masses into the substellar regime. 

\end{abstract}




\section{Introduction}
\label{Intro}
Ultracool dwarfs (UCDs) are ubiquitous cool objects with an effective temperature ($T_{\mathrm{eff}}$) below $\sim$ 2700\,K, including the late M-dwarfs, and extend to the sub-stellar Brown Dwarf (BD) regime including L, T, and Y dwarfs \citep{kirkpatrick_1997, Rajpurohit_2013}. 

Independent measurements of the mass, age, \text{\teff}, and surface gravity of BDs are vital for testing the current formation and evolutionary substellar models {\citep{ 2008ApJ...689.1327S, 2011ApJ...736...47B, 2015A&A...577A..42B, 2020A&A...637A..38P, 2021ApJ...920...85M}}. \text{\teff} and surface gravity can be estimated using photometry and spectroscopic analyses, however, as BDs continuously cool with age, these parameters are not translatable to a precise mass. This results in an age-mass degeneracy for a given luminosity and temperature \citep{1997ApJ...491..856B}, with lower mass younger BDs having similar \text{\teff} and luminosity to higher mass, older counterparts, thus making field BDs extremely difficult to characterise. The uncertainties in these measurements, particularly in mass and age, pose significant difficulties for their use in testing and refining evolutionary models and brown dwarf cooling models \citep{1989ApJ...345..939B}.

UCDs in multiple systems, particularly those with brighter companions, play an important role in disentangling mass-age-luminosity degeneracy. These ``benchmark'' systems provide model-independent constraints for the UCD components by leveraging the fundamental parameters of the primary (brighter) companion \citep{2006MNRAS.368.1281P}. Significant progress has been made in identifying UCD benchmark systems \citep{2000ApJ...531L..57B, 2009MNRAS.395.1237B, Burningham_tdwarf, 2010AJ....139..176F, 2010MNRAS.404.1817Z, 2011MNRAS.410..705D, 2012ApJ...760..152L(luwe19), 2012MNRAS.422.1922P, 2013MNRAS.431.2745G, 2020MNRAS.499.5302D, 2024AJ....167..253R, 2024MNRAS.533.3784B}, providing the necessary data to probe UCD formation scenarios. UCD benchmarks also assist in probing low-mass binary properties (e.g., frequency, orbital separation, and mass-ratio distribution), as they are invariant to the initial kinematic and cloud structure conditions of the gas in the star-forming regions in which they are formed \citep{2009MNRAS.397..232B,2011IAUS..270..133B, 2012MNRAS.419.3115B}, thus enabling an unbiased study of their distributions.

One approach to addressing the mass-age-luminosity degeneracy for field brown dwarfs is through the age dating of benchmark UCD systems. Assuming that the system is coeval and formed in the same molecular cloud, we can consider the age and composition of the primary to be the same as those of the companion. Although traditional stellar-age dating methods are not directly applicable to brown dwarfs, age estimates can be achieved using coeval clusters or associations, particularly in young clusters (t$<$10 Myr) \citep{2018ApJ...856...23G, 2020MNRAS.492.5811B}, where brown dwarfs are most luminous in their early ages.  Gyrochronology, which leverages the relationship between rotational evolution and age, offers another valuable tool for age dating UCDs, particularly for those lacking direct mass measurements. This method is particularly useful for binaries and hierarchical systems, where the rotation rates of primary stars or companions can calibrate the ages of the UCDs, yielding an empirical
stellar ages with a precision of 15$\%$–20$\%$ \citep{2019ApJ...871...39M}. Another method involves directly measuring dynamical masses in tight binaries using radial velocities or astrometric orbits. These measurements can constrain the orbital masses to better than 10$\%$ precision \citep{2017ApJS..231...15D, 2023MNRAS.519.1688D}, constraining age estimates to 10-20$\%$ uncertainties on evolutionary tracks compared with the 50–100$\%$ uncertainties typical of main-sequence stellar dating \citep{2009IAUS..258..317B}.

In this paper we present ULAS J125015.55+045506.6AB (hereafter J1250+0455AB), previously catalogued as a single M9 ultracool dwarf by \citet{2025MNRAS.538.3144C} but now resolved into a close binary of two early‐L dwarfs separated by $12.2\pm1.5$\,AU. J1250+0455AB is itself part of a hierarchical triple with the early‐M dwarf Gaia DR3 3705763723623026304 (hereafter J1250+04553; \citealt{2024MNRAS.533.3784B}). \citet{2024NatAs...8..223L} estimate an age of $\sim0.8,$Gyr and a slightly sub‐solar metallicity ([Fe/H] = $-0.168$) for J1250+04553, in agreement with values reported by large‐scale spectral surveys \citep{2022yCat..22600045D,2024A&A...684A..29V}.

We resolved J1250+0455AB using the Large Binocular Telescope (LBT), specifically the LBT Utility Camera in the Infrared (LUCI), a NIR imager with adaptive optic capabilities, and subsequently characterise both components of J1250+0455AB in our analysis. 
The structure of the paper is as follows: Section \ref{target selection} outlines the identification of our target; Section \ref{observations} provides details of the observations conducted with LBT/LUCI; Section \ref{Analysis} describes the analysis of J1250+0455AB; and Section \ref{conclusion} presents our conclusions and final remarks.

\section{Target Selection}
\label{target selection}
J1250+0455AB was selected from the UCD companion catalogue of \citet{2024MNRAS.533.3784B}. To identify systems likely to harbour unresolved binarity in the Gaia data, we applied three diagnostic criteria for a partially resolved binary, motivated by the work of \cite{2023A&A...674A...9H} using the following criteria for $\gaia$ parameters:

\begin{itemize}
    \item \textbf{{ipd\_gof\_harmonic\_amplitude (IPDgofha)} $>$ 0.1}: The criterion delineates the goodness of fit as a function of the position angle of the scan direction. This parameter assists in measuring the significance of the scanning angle on the Point Spread Function (PSF) fitting process, indicating elongations and the presence of unresolved binarity. 

        \item \textbf{RUWE $>$ 1.4}: The Renormalised Unit Weight Error (RUWE) serves as a reliable metric to assess the goodness-of-fit for the astrometric solution of a $\gaia$ source. As detailed in \cite{EDR3_astrometry} a RUWE $\le$ 1.4 is indicative of a well-behaved single-star solution. It has been shown that RUWE values significantly above this threshold can be attributed to photocentric motions of unresolved objects \citep{2021ApJ...907L..33S}, such as astrometric binaries, which are not revealed by the $\gaia$ Image Parameter Determination (IPD) statistics, and therefore complement the IPD in binary detection.

    \item \textbf{$G - G_{\mathrm{RP}} > 3\sigma\bigl({G-G_{\mathrm{RP}}}(\text{SpT})\bigr)$}:  
    When multiple components remain unresolved by \textit{Gaia}, they are recorded as a single source, thereby inflating the measured brightness, producing a colour mismatch relative to that expected for a single star of the same spectral type. In \emph{Gaia}~DR3 \citep{2023A&A...674A...1G}, there are no pipeline corrections for separating the flux from such genuinely merged detections \citep{2023A&A...674A...2D}, resulting in blending of photometry from both sources. To identify potential unresolved pairs, we used photometric relations from \cite{2019AJ....157..231K} to estimate the typical $\g$–$\grp$ values for the spectral types of our UCD targets, as \cite{2019AJ....157..231K} found that it had the tightest relation to the spectral type and compared this to our targets measured $\g$ - $\grp$. We selected any source that deviated by more than 3$\sigma$ from the mean $\g$–$\grp$ of the UCDs spectral type. 
\\    
We did not adopt the blending fraction  ($\beta$) from \citet{2021A&A...649A...3R} because such fraction-based methods mainly apply when a secondary source is at least partially resolved, which is not the case for the compact systems considered here.

\end{itemize}
\cite{2023A&A...674A...9H} also adopted the threshold \textbf{ipd\_frac\_multi\_peak (IPDfmp)} $\leq 2$, which reflects the proportion of scans that exhibit dual peaks and suggests possible secondary sources. However, our selection criteria do not include this measure, as dual peaks may not always be evident depending on the scan orientation or close proximity of the components.
\\

J1250+0455AB satisfied all criteria and was consequently prioritised for high‐resolution LUCI imaging. Other systems from the initial catalogue did not meet one or more requirements or could not be observed because of guide‐star limitations and are not discussed further here.

\begin{table}
\caption{Selected properties of J1250+0455AB.}
\label{J1250+0455_properties}
\begin{tabular*}{\columnwidth}{@{\extracolsep{\fill}}lll@{}}
\toprule
Parameter & Unit & Value \\ \midrule
\multicolumn{3}{c}{\textbf{Information}} \\ \midrule
$\gaia$ ID & -- & 3705763723623660416 \\
RA & $^{\circ}$ & 192.6 \\
DEC & $^{\circ}$ & 04.91 \\ \midrule
\multicolumn{3}{c}{\textbf{Photometry}} \\ \midrule
$\gaia$ G & mag & 20.46 $\pm$ 0.01$^{1}$ \\
$\gaia$ Grp & mag & 18.27 $\pm$ 0.02$^{1}$ \\
$\gaia$ Gbp & mag & 21.49 $\pm$ 0.16$^{1}$ \\
2MASS J & mag & 15.15 $\pm$ 0.06$^{2}$ \\
2MASS H & mag & 14.34 $\pm$ 0.06$^{2}$ \\
2MASS K$\mathrm{_s}$ & mag & 13.80 $\pm$ 0.05$^{2}$ \\
UKIDSS H & mag & 14.42 $\pm$ 0.01$^{3}$ \\
UKIDSS K & mag & 13.85 $\pm$ 0.01$^{3}$ \\ \midrule
\multicolumn{3}{c}{\textbf{Astrometry}} \\ \midrule
Parallax & mas & 13.93 $\pm$ 1.13$^{1}$ \\
pmRA & mas\,yr$^{-1}$ & $-129.4 \pm 1.39^{1}$ \\
pmDEC & mas\,yr$^{-1}$ & $42.47 \pm 1.34^{1}$ \\
\texttt{ipd\_gof\_harmonic\_amplitude} & -- & 0.15$^{1}$ \\
RUWE & -- & 1.51$^{1}$ \\ \midrule
\multicolumn{3}{c}{\textbf{Astrophysical Parameters}} \\ \midrule
SpT & -- & M9$^{4}$ \\
$T_{\rm eff}$ & K & $\sim$2420$^{5}$ \\
Mass & $\mathrm{M}_\odot$ & $\sim$80$^{5}$ \\ \bottomrule
\end{tabular*}
\smallskip
\footnotesize
$^{1}$ \gaia\ DR3 \citep{2023A&A...674A...1G};
$^{2}$ 2MASS \citep{Two_Mass};
$^{3}$ UKIDSS \citep{2007MNRAS.379.1599L};
$^{4}$ \cite{2025MNRAS.538.3144C};
$^{5}$ Relations from \cite{2013ApJS..208....9P}.
\end{table}

\begin{figure*}
\begin{center}
\includegraphics[clip,width=1.5\columnwidth]{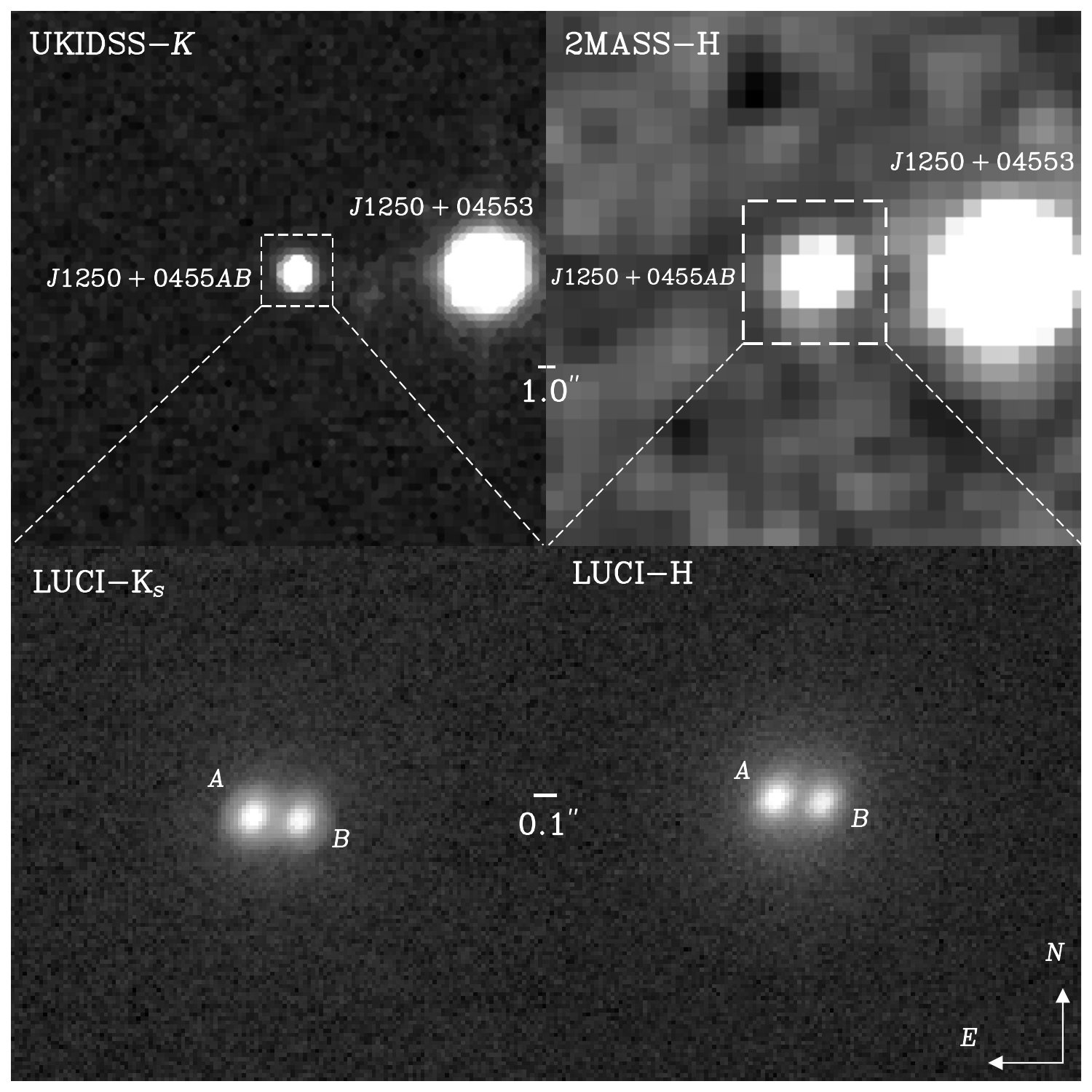}

\caption{
\textbf{[Top panel]}: Images of the system J1250+0455AB and its nearby companion J1250+04553 in the UKIDSS-K (left) and 2MASS-H (right) bands. The white dashed squares highlight the unresolved PSF of J1250+0455AB in both the images. The dashed lines connect these squares to the corresponding LUCI-K$\rm{_s}$ and LUCI-H images where the individual components of the binary system are resolved. The UKIDSS-K and 2MASS-H images cover a field of view of 0.5$\arcmin$ × 0.5$\arcmin$. The 1$\arcsec$ scale shown is accurate only for the UKIDSS and 2MASS images. \textbf{[Bottom panel]}: Resolved images of the J1250+0455AB system in the LUCI-K$\rm{_s}$(left) and LUCI-H (right) bands. The LUCI images cover a smaller field of view of 2$\arcsec$ × 2$\arcsec$. The 0.1$\arcsec$ scale shown is accurate for the LUCI images. The primary (A) and secondary (B) components of the system are labelled in both LUCI images. In all images, the North is oriented upwards, and the East is to the left. }
\label{fig1}
\end{center}
\end{figure*}

\section{Observations and data reduction}
\label{observations}
\subsection{Observations with LBT/LUCI}

Using the Large Binocular Telescope (LBT) of the Mount Graham International Observatory, specifically the LBT Utility Camera in the Infrared  \citep[LUCI,][]{2018SPIE10702E..2XP}, in Single Conjugate Adaptive Optics
(SCAO) mode with the second-generation AO instrument Single Conjugated Adaptive Optics Upgrade for LBT \citep[SOUL, ][]{2016SPIE.9909E..3VP, 2021arXiv210107091P}, J1250+0455 was observed. A summary of selected properties of J1250+0455 are presented in Table \ref{J1250+0455_properties}. 

Observations of J1250+0455 were made with LUCI1, located on the left-hand side (SX) of the LBT, under program ID 2028203, observed on the 15th February 15, 2024. The N30 camera was used in the \texttt{ADAPTIVE} mode with a field of view of 30$\arcsec$ x 30$\arcsec$ and a pixel scale of 0.015$\arcsec$/pix. A total of 15 frames were observed in both the H (1.653 $\mu$m) and K$\rm{_s}$ (2.153 $\mu$m) photometric bands, with each band receiving a total exposure time of 75s, through a 5s exposures for each frame. In order to allow for an efficient reduction of the sky background, each image was dithered according to a rectangular pattern with a maximum displacement of 5$\arcsec$ in either the X- or Y-direction. LUCI necessitates an AO reference star, which can be either on-axis (R $<$ 16.5\,mag) or off-axis. J1250+0455 was too faint to be used as an on-axis guide star; thus, we selected J1250+04553, located 10.44$\arcsec$ away, as the off-axis guide star, as it is the only star sufficiently bright and within the 1\,$\arcmin$ field of view of LUCI1 to enable a feasible observation. 
All observations of J1250+0455 were obtained within an airmass of 1.176-1.202. During the observations, the Differential Image Motion Monitor (DIMM) was operational only briefly while observing the first source. However, based on the measurements obtained prior to and following the observation period, the seeing conditions along the telescope line of sight were estimated to range from $1.0~\arcsec$ to $1.3~\arcsec$. We conducted observations in the H and K$\rm{_s}$ bands, where our targets were the brightest. The secondary companion, being cooler than the primary, has greater brightness at near-infrared wavelengths; thus, we opted for the K$\rm{_s}$band rather than the J band to maximise the signal-to-noise of the image, improving our ability to identify subtle variations in the PSF morphology. The full details of the observational conditions for each frame in both photometric bands are listed in Table \ref{binary_profile_table}. \\
From the AO telemetry data, we estimated a variable Strehl Ratio (SR) between 20$\%$ and 30$\%$ at 1650\,nm with a Full Width at Half Maximum (FWHM) of 50\,mas in the AO reference star. Considering a typical vertical distribution of the seeing and the 10.44$\arcsec$ distance, we expect to have on the scientific target an FWHM of approximately 60\,mas in the H band and between 60-70\,mas in Ks. Considering the variability of the seeing vertical distribution, the measured values were consistent with the AO telemetry estimations.

\subsection{Data reduction}
Calibration and sky subtraction of the individual exposures were performed using a dedicated pipeline developed at INAF - Rome Observatory to reduce LBT LUCI imaging data \citep{2014A&A...570A..11F}.
Initially, the reduction procedure involved the removal of the dark current and flat field correction. To remove the dark current, a median stacked dark image (masterdark) that is obtained by combining a collection of dark frames using the same detector integration time (DIT) and number of DITs (NDIT) as the scientific dataset is subtracted from both the scientific and sky images. The dark-subtracted images were then divided by a median stacked flat image  (masterflat) acquired from a combination of a set of dark-subtracted flat images. The pre-reduced images were then sky-subtracted using a mean sky image taken at the beginning and end of the observation sequence. The final image registration and coaddition were performed using the \texttt{IMCOMBINE} task within the IRAF \citep{1986SPIE..627..733T,1993ASPC...52..173T}.

J1250+0455AB has been observed in previous near-infrared surveys, including the UKIRT Infrared Deep Sky Survey (UKIDSS; \citealt{2007MNRAS.379.1599L}) and Two Micron All Sky Survey (2MASS; \citealt{Two_Mass}) . However, the resolution of these surveys was insufficient to resolve the individual components of J1250+0455AB, necessitating the use of AO to image the system. We present both the UKIDSS-K and 2MASS-H images, along with the higher-resolution LUCI H and K$\rm{_s}$ band images of J1250+0455 in Fig.\ref{fig1}. In both the LUCI bands, J1250+0455 was clearly resolved as a binary system with two components. The components are similar in brightness, with the eastern component (J1250+0455A) exhibiting a slightly larger flux than the western component (J1250+0455B), suggesting a near-equal mass system. Our analysis designates the brighter component, J1250+0455A, as the primary.

\section{Analysis}
\label{Analysis}
\subsection{Binary parameters}
\label{binary parameters}
\subsubsection{Separation $\&$ Position Angle}
We measured the coordinates on the CCD detector of each source in the stacked H and K$\rm{_s}$ images using the \texttt{PHOTUTILS} \texttt{DAOStarFinder} class in \texttt{PYTHON} \citep{1987PASP...99..191S}, and from these positions computed the separation between their centroids.

The \texttt{DAOSTARFINDER} algorithm identifies centroids by fitting the marginal x- and y-dimensional distributions with a Gaussian kernel to those of the unconvolved input image. Following the identification of the centroids, the separation was computed in arcseconds by converting the pixel distance using the pixel scale of LUCI (0.015$\arcsec$/pix). The calculated separations for each science image across both photometric bands are presented in Table \ref{binary_profile_table}. We opted to use the separation derived from the stacked K$\mathrm{_s}$-band image, which has a superior SNR compared to the stacked H-band image. Specifically, the K$\mathrm{_s}$-band image provided SNRs of 159 and 124 for J1250+0455A and J1250+0455B, respectively, whereas the H-band image yielded lower SNRs of 99 and 75, respectively. This resulted in a separation of 170 $\pm$ 15\,mas between the resolved components.  
The FWHM of the PSF was determined by fitting a composite model comprising a Gaussian profile for the central core and Moffat profile for the extended halo. The Gaussian component performed better near the diffraction limit, whereas the Moffat component accounted for the residual flux in the outer halo. The resulting FWHM values were $0.076 \pm 0.006\arcsec$ and $0.078 \pm 0.006\arcsec$ in  K$\rm{_s}$ and $0.083 \pm 0.007\arcsec$ and $0.085 \pm 0.008\arcsec$ in H for J1250+0455A and J1250+0455B components, respectively.

To determine the Position Angle (PA), pixel coordinates were first transformed into World Coordinate System (WCS) coordinates using the \texttt{ASTROPY} \citep{2013A&A...558A..33A} WCS class. Subsequently, the PA was calculated using the \texttt{PYASTRONOMY} \citep{pya} \texttt{pysal} module. The stacked K$\mathrm{_s}$ image provided a PA of 85.84 $\pm$ 0.2 $^\circ$, and the PA for each science image is presented in Table \ref{binary_profile_table}.

\subsubsection{Flux ratio $\&$ Photometry}
\label{fr and phot sec}
To derive the flux ratio between the primary and secondary components of J1250+0455AB aperture photometry was performed using the \texttt{PHOTUTILS} \texttt{ApertureStats} class on the co-added images, with an aperture radius of 0.0675$\arcsec$ for both components. Aperture photometry was used with non‐overlapping circular apertures that fully isolate each component. The background levels and uncertainties for each frame were estimated using an annular region with a width of 0.0675$\arcsec$ around each source. The annulus was configured to have a 0.015$\arcsec$ gap from the edge of the aperture used for the flux measurements. 
The flux ratio was then calculated as the ratio between the primary and secondary fluxes, yielding 1.27 $\pm$ 0.01 in K$\rm{_s}$, with the flux ratio for all images in both bands, shown in Table \ref{binary_profile_table}.

\begin{figure}
  \centering
  \includegraphics[width=\columnwidth]{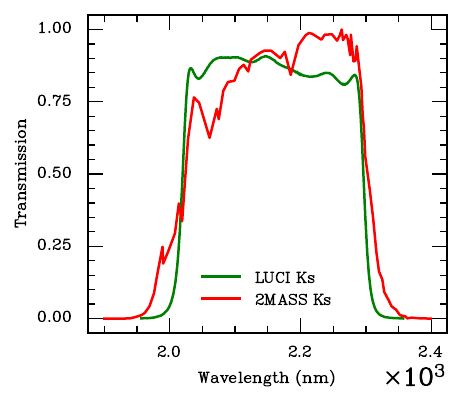}
  \caption{%
    Absolute magnitude $M_{K_s}$ as a function of spectral type.  
    The black line represents the polynomial fit from 
    \citet{2012ApJS..201...19D}, whereas the blue points denote selected 
    samples from the same catalogue.  
    The blue and red dashed lines indicate the $M_{K_s}$ values for 
    J1250+0455A and J1250+0455B, respectively, with the shaded regions 
    reflecting the corresponding uncertainties.%
  }
  \label{fig:2mass_transmission}
\end{figure}

Calibration of our LUCI K$\rm{_s}$ photometry was performed against the 2MASS K$\rm{_s}$ system, chosen because among all publicly available survey filters from the Spanish VO Filter Profile Service \citep{2012ivoa.rept.1015R, 2020sea..confE.182R, 2024A&A...689A..93R} its bandpass most closely matches that of LUCI, as shown in Fig.~\ref{fig:2mass_transmission}.  We interpolated the 2MASS transmission curve onto the LUCI wavelength grid and assessed their agreement by computing the sum of squared differences, finding deviations of only a few percent across the full bandpass. We therefore used the known unresolved 2MASS K$\rm{_s}$ magnitude of our target system to determine the photometric zero point and applied that zero point to the individual LUCI fluxes of the primary and secondary components to recover their 2MASS‐scale magnitudes.

\begin{equation}
    \mathrm{ZP}_{\mathrm{H, Ks}} = \mathrm{2MASS}_{\mathrm{H, Ks}} - 2.5\mathrm{log}_{10}(\mathrm{F}_{\mathrm{total, H, Ks}})
\end{equation}

where $\text{ZP}_{\text{H, Ks}}$ are the photometric zero-points in the two bands, $\text{2MASS}_{\text{H, Ks}}$ is the unresolved 2MASS magnitudes of J1250+0455AB, and $\mathrm{F}_{\mathrm{total, H, Ks}}$ is the combined flux of both the primary and secondary components from the aperture photometry. 

To estimate the individual 2MASS magnitudes for the components, we used the following: 

\begin{equation}
\text{2MASS}_{\mathrm{H, Ks}}^{\mathrm{A, B}} = \mathrm{ZP}_{\mathrm{H, Ks}} - 2.5 \log_{10}(F_{\mathrm{H, Ks}}^{\mathrm{A, B}})
\end{equation}

where $\text{2MASS}_{\mathrm{H, Ks}}^{\mathrm{A, B}}$ represents the \textbf{measured} 2MASS magnitudes for the primary and secondary components in the H and  K$\rm{_s}$ bands and $F_{\mathrm{H, Ks}}^{\mathrm{A, B}}$ denotes the flux of the primary and secondary components in the corresponding LUCI bands. The estimated 2MASS magnitudes are listed in Table \ref{J1250+0455_binary_properties}. 

\begin{figure}
\begin{center}
\includegraphics[clip,width=\columnwidth]{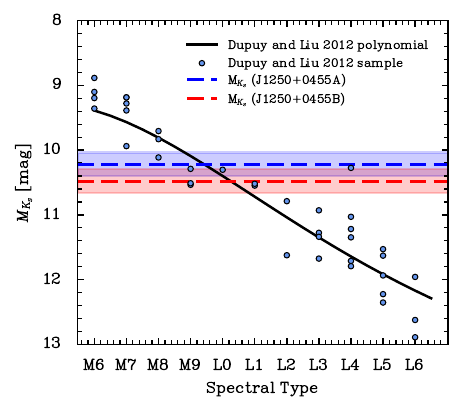}
\caption{Absolute magnitude M$\mathrm{_{K_s}}$ as a function of spectral type. The black line represents the polynomial fit from \protect\cite{2012ApJS..201...19D}, whereas the blue points denote selected samples from the same catalogue. The blue and red dashed lines indicate the M$\mathrm{_{K_s}}$ values for J1250+0455A and J1250+0455B, respectively, with the shaded regions reflecting the corresponding uncertainties.}

\label{sptphot_ks}
\end{center}
\end{figure}

Fig.\ref{sptphot_ks} presents the absolute magnitude (M$\mathrm{_{K_s}}$) as a function of spectral type for a selection of M and L dwarfs, based on the compilation by \cite{2012ApJS..201...19D} along with the derived M$\mathrm{_{K_s}}$ values for the individual components of J1250+0455AB. The spectral types of components A and B were consistent within their respective uncertainties. Specifically, we derived a spectral type of M9 $\pm$ 0.6 for J1250+0455A, and L0 $\pm$ 0.6, for J1250+0455B. The dominant source of uncertainty in M$\mathrm{_{K_s}}$ arises from the parallax, as the $\gaia$ DR3 parallax measurement of 13.93 $\pm$ 1.13\,mas has a large associated uncertainty. This significant uncertainty is due to the wobbling of the photocentre during Gaia observations, leading to a poor fit of the astrometric solution \citep{2024A&A...688A...1C}. The uncertainty of each spectral type was derived by propagating all parameter uncertainties.

To derive the bolometric luminosity (\(\log \left(\mathrm{L} / \mathrm{L}_{\odot}\right)\)), we applied the 4th order polynomial relations for bolometric corrections from \cite{2023AAS...24120311S}, specifically utilising the \(\text{BC}_{\text{K}_{s \text{FLD}}}\) corrections. Following the approach outlined in \cite{2024MNRAS.533.3784B}, we compute \(\log \left(\mathrm{L} / \mathrm{L}_{\odot}\right)\) values, incorporating uncertainties from both the intrinsic scatter in the polynomial relation and the error in M\(_\text{K}{_\text{s}}\). This resulted in bolometric luminosities of \(\log \left(\mathrm{L} / \mathrm{L}_{\odot}\right) = -3.45 \pm 0.04\) dex and\(\log \left(\mathrm{L} / \mathrm{L}_{\odot}\right) = -3.58 \pm 0.04\) dex for components A and B, respectively.

\begin{table}
\caption{Measurements of binary separation, position angle and flux ratios for J1250+0455AB for each of the 15 science images in both H-band and K$\rm{_s}$-band with the final row giving the values derived from the stacked image.} 
\label{binary_profile_table}
\begin{tabular}{ccccccccc}
\hline
\multicolumn{3}{c}{H band} & \multicolumn{3}{c}{K$\rm{_s}$ band} \\ \hline
Sep (mas) & PA ($^\circ$) & Flux ratio & Sep (mas) & PA ($^\circ$) & Flux ratio \\ \hline
171.4  & 84.97 & 1.28 $\pm$ 0.02 & 169.4 & 86.56 & 1.29 $\pm$ 0.01 \\
173.9  & 86.19 & 1.31 $\pm$ 0.01 & 167.6 & 86.13 & 1.28 $\pm$ 0.02 \\
164.1  & 83.47 & 1.27 $\pm$ 0.03 & 172.7 & 86.31 & 1.28 $\pm$ 0.01 \\
164.3  & 85.06 & 1.24 $\pm$ 0.03 & 172.3 & 85.79 & 1.29 $\pm$ 0.01 \\
170.0  & 85.15 & 1.30 $\pm$ 0.03 & 167.4 & 86.36 & 1.26 $\pm$ 0.01 \\
168.5  & 85.74 & 1.26 $\pm$ 0.03 & 169.9 & 86.25 & 1.28 $\pm$ 0.01 \\
176.8  & 86.99 & 1.26 $\pm$ 0.02 & 170.9 & 85.65 & 1.30 $\pm$ 0.01 \\
173.9  & 86.26 & 1.28 $\pm$ 0.01 & 171.6 & 85.64 & 1.27 $\pm$ 0.01 \\
169.7  & 85.77 & 1.25 $\pm$ 0.03 & 167.5 & 85.87 & 1.28 $\pm$ 0.02 \\
170.4  & 86.78 & 1.26 $\pm$ 0.02 & 172.3 & 85.84 & 1.25 $\pm$ 0.01 \\
173.8  & 87.36 & 1.25 $\pm$ 0.02 & 167.0 & 84.95 & 1.26 $\pm$ 0.02 \\
171.3  & 85.85 & 1.28 $\pm$ 0.03 & 170.8 & 85.89 & 1.29 $\pm$ 0.01 \\
168.3  & 86.34 & 1.26 $\pm$ 0.02 & 164.2 & 85.26 & 1.25 $\pm$ 0.01 \\
170.2  & 85.93 & 1.28 $\pm$ 0.01 & 168.3 & 86.13 & 1.24 $\pm$ 0.02 \\
171.4  & 85.62 & 1.27 $\pm$ 0.02 & 170.3 & 85.03 & 1.26 $\pm$ 0.02 \\
\hline
170.5 & 85.83 & 1.27 $\pm$ 0.01 & 169.5 & 85.9 & 1.27 $\pm$ 0.01 \\
\hline
\end{tabular}
\end{table}

\subsubsection{Combined spectral fitting}
\label{spectral fitting}

We aimed to estimate the spectral types of the resolved components of J1250+0455AB using a combined spectral analysis approach. \cite{2025MNRAS.538.3144C} initially classified the unresolved system as an M9 dwarf based on low-resolution near-infrared (NIR) spectroscopy obtained using the SpeX instrument at the NASA Infrared Telescope Facility (IRTF) \citep{SpeX}. Resolving the binary components provides an opportunity to refine this classification.

To estimate the spectral types of the individual components, we utilised empirical spectral templates from the SpeX Prism Library Analysis Toolkit (SPLAT) \citep{burgasser2014}. These templates include standard spectra from the IRTF’s Short-wavelength Cross-Dispersed (SXD) mode, which provides high-quality empirical standards across M- and L-dwarf spectral types.

As noted in Sec.~\ref{Intro}, available constraints point to a system age of ~0.8 Gyr and slightly sub-solar metallicity ([Fe/H] $\approx -0.17$
); we revisit the age diagnostics and confirm this estimate in Sec.~\ref{age} below. Since empirical templates that jointly reflect modest metal deficiency at intermediate gravities are lacking, we bracketed the component spectra using INT-G standards (M9–L2) alongside the available low-metallicity/subdwarf sequences.
Because empirical templates that jointly reflect modest metal deficiency at field gravities are lacking, we bracketed the component spectra using INT-G standards (M9–L2) alongside the available low-metallicity/subdwarf sequences.
Given the absence of benchmarks that simultaneously exhibit mild metal-poor ([Fe/H]\,$\approx$\,-0.2 to –0.4) and young (INT-G) characteristics, and noting that the only available low-metallicity templates are limited to subdwarf standards, we confined our fits to intermediate-gravity (INT-G) templates spanning spectral types M9 to L2, including half-subtypes.

\begin{figure*}
\begin{center}
\includegraphics[]{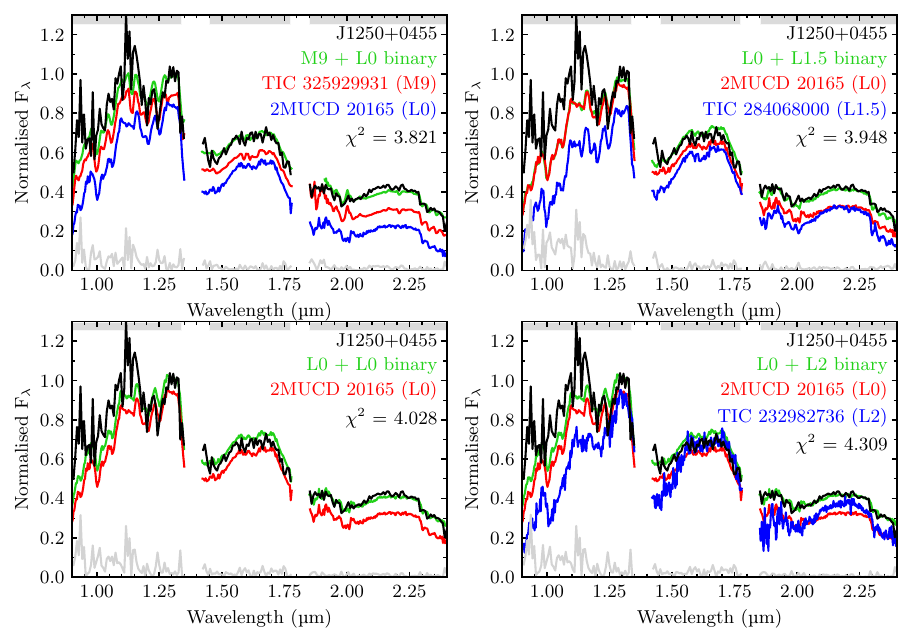}
\caption{ Best-fitting binary templates for J1250+0455AB are shown. The black line represents the observed spectrum of J1250+0455AB \citep{2025MNRAS.538.3144C} with the binary fits in green. The primary and secondary components of the fit are marked in red and blue, respectively. The four best fits, ranked by $\chi^2$, are arranged with the best fit at the top left. Grey lines indicate the residuals between the observed (black) and best-fit binary (green) spectra. Grey bars at the top denote fitted spectral regions, with water absorption gaps.}
\label{spectral_fit_plot}
\end{center}
\end{figure*}

We fitted multiple binary combinations as shown in Fig.\,\ref{spectral_fit_plot}, displaying the best template fits. The best fitting solution is an M9\,$\pm$\,1.1 primary combined with an L0\,$\pm$\,1.4 secondary, both classified as INT-G.

This estimate is consistent with our photometric classification of the individual stars, which suggests that the system contains two nearly equal-mass M/L boundary dwarfs. The uncertainties for each component of the binary fits were calculated separately by fixing the secondary component at the best-fitting spectral type, which allowed us to calculate the chi-squared values for a range of primary spectral types. We then fitted a polynomial to these chi-squared values and defined the uncertainty on the primary component to be the difference in spectral types between the best-fitting primary type (i.e.\ the spectral type with the lowest chi-squared value, $\chi^2_{\min}$) and that corresponding to a chi-squared value equal to $\chi^2_{\min} + 1$ (corresponding to a 1$\sigma$ uncertainty). We then repeated this process by fixing the primary component at its best-fitting spectral type to calculate the uncertainties for the secondary component independently, resulting in independent uncertainties for each component of the binary spectral fits.

\subsection{Physical parameters}
\subsubsection{Age}
\label{age}
Accurate age determination remains an elusive goal for many stellar types, particularly for M dwarfs. These stars undergo hydrogen fusion at significantly slower rates than their more massive counterparts, resulting in main-sequence lifetimes far exceeding the current age of the universe \citep{1997ApJ...482..420L}. Consequently, traditional age dating methods, such as isochrones and astroseismology, which are effective for massive stars, often prove unreliable for M dwarfs because of their long stable main-sequence lifetime, resulting in extremely slow changes in temperature and luminosity \citep{2010ARA&A..48..581S, 2015A&A...577A..42B}.

Observations of M dwarf magnetic activity have elucidated a correlation between stellar age and rotational period, offering a viable method for dating these stars \citep{2015ApJ...812....3W}. The fundamental principle underlying this approach is that stars ``spindown'' -  exhibiting progressively longer rotational periods as they age, primarily due to angular momentum loss via magnetised stellar winds \citep{1962AnAp...25...18S, 1967ApJ...148..217W}. The seminal work of \cite{1972ApJ...171..565S} introduced a power-law relationship between the stellar rotation and age, laying the groundwork for gyrochronology \citep{2003ApJ...586..464B, 2007ApJ...669.1167B}, relating the stellar age to the rotational period ($P_{\rm{rot}}$) and the effective temperature ($T_{\mathrm{eff}}$). The pursuit of calibrating an empirical gyrochronology relation has been assisted by large photometric surveys including Kepler \citep{2010Sci...327..977B}, Transiting Exoplanet Survey Satellite (TESS) \citep{2015JATIS...1a4003R}, K2 \citep{2014PASP..126..398H}, MEarth \citep{2012AJ....144..145B}, and Zwicky Transient Facility (ZTF) \citep{2019PASP..131a8003M,2022ASPC..532..277M}. These surveys provided the necessary data to determine stellar rotation from their light curves \citep{2014A&A...572A..34G, 2021AJ....161..189L,2021ApJ...913...70G, 2023ApJ...954L..50E,2024AJ....167..159L}, which reveals a more complicated substructure of periodicity than that first proposed by \cite{1972ApJ...171..565S} \citep[see][and references therein for further details]{2024AJ....167..159L}. Thus, recent gyrochronology relations leverage empirical calibrations with benchmark stars with ages determined from asteroseismology, open clusters White Dwarf cooling ages, and wide binary systems. 

\begin{figure}
\begin{center}
\includegraphics[clip,width=\columnwidth]{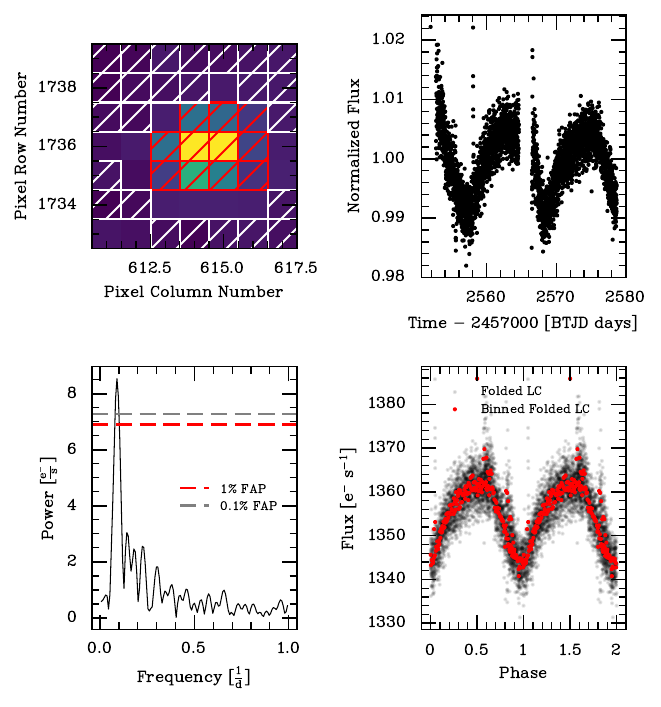}
\caption{[\textbf{Top left]}: TESS image of J1250+04553, featuring our aperture mask in red and the background marked in white. \textbf{[Top right]}: Background-subtracted light curve of J1250+04553. \textbf{[Bottom left]}: Lomb-Scargle periodogram of J1250+04553, highlighting the strongest signal at 
11.48 days, surpassing the 1$\%$ and 0.1$\%$ false-alarm probability (FAP) thresholds. \textbf{[Bottom right]}: Phase-folded light curve of J1250+04553, using the best period identified from the Lomb-Scargle analysis, with the data binned every 2 hours and displayed in red.}
\label{period_plot}
\end{center}
\end{figure}

Bright Main-Sequence stars, when part of a wide binary system with a UCD, are instrumental in characterising their companions. Assuming coevality, the astrophysical parameters of the primary star, including its metallicity and age, parameters typically challenging to measure for low-mass stars, can be applied to their companion. Independent age estimates are especially valuable as they are essential for disentangling the mass-age-observable degeneracy associated with the brown dwarf companion. \citet{2024MNRAS.533.3784B} first identified J1250+04553, an M2.5V star, as a wide binary companion to J1250+0455AB, located at an angular separation of 10.44$\arcsec$. We aimed to determine the rotational period as the basis for the age calculation. We searched the Mikulski Archive for Space Telescopes (MAST) and found TESS observations under TIC 380727853 relevant to our target. To refine our analysis, we developed a custom TESS light curve by examining the Target Pixel File (TPF) J1250+04553. The construction of our own light curves is premised on the possible contamination of J1250+0455AB in the light curve because of its proximity and the minimisation of noise by masking any pixel dominated by the background. Utilising the \texttt{lightkurve} software \citep{2020AAS...23540904B}, we identified sector 46 containing observations of the source, offering a 10-minute cadence. The flux contribution was gauged from selected pixels exhibiting values exceeding 12$\sigma$ above the median flux of the image, centred on the reference pixel of the target. These pixels were then aggregated to compute the raw flux of the target. The background flux was ascertained by summing pixels registering above 0.1$\sigma$ of the median flux, which was then subtracted from the raw flux to yield a background-corrected light curve. We also conducted a parallel period search in archival ZTF light curves (object ID: 474210300004550) in the $g$, $r$, and $i$ bands; however, no significant periodic signals were detected. Subsequent analysis of the corrected light curve via a Lomb-Scargle periodogram revealed a pronounced peak at a period of 11.48 days, significantly above the 99.9$\%$ False Alarm Probability (FAP) threshold. Folding the light curve at this detected period exhibited distinct periodic variability, further accentuated by binning the light curve at 2-hour intervals, as shown in Fig. \ref{period_plot}. By using the gyrokinematic relations developed in \cite{2024AJ....167..159L} we input the rotational period, effective temperature from $\gaia$ and absolute $\gaia$ magnitude to derive an age of $0.58^{+0.07}_{-0.06}$ Gyr. We note that the quoted uncertainties only capture the internal scatter of the model fit and 
do not account for observational errors in $T_{\mathrm{eff}}$ or $P_{\mathrm{rot}}$; thus, they are likely to be underestimated.

We note that an age of 0.81\,Gyr was obtained by \citet{2024NatAs...8..223L} for J1250+04553. While this is broadly consistent with our derived age, slight discrepancies in both  $P_{\mathrm{rot}}$ and $T_\mathrm{eff}$ explain the difference in determined age. In terms of metallicity, \citet{2024NatAs...8..223L} also report a sub-solar metallicity of [Fe/H] = $-0.168$. This is in agreement with the findings of recent large-scale spectral fitting catalogues, which have derived sub-solar metallicities for this star, including [Fe/H] = $-0.37$ from \citet{2022yCat..22600045D} and [Fe/H] = $-0.50$ from \citet{2024A&A...684A..29V}. Although there is some variation in the metallicity estimates, all are consistent with a sub-solar value.

\label{Age}

\subsubsection{Mass, Teff and Radii}
\label{mass_teff_radii section}
To estimate the masses, effective temperatures, and radii of both components of J1250+0455AB, we utilise two evolutionary models: the models of \citet{2015A&A...577A..42B} (hereafter BCAH15) and the “hybrid” models of \citet{2008ApJ...689.1327S} (hereafter SM08), both of which assume solar metallicity. These models were selected because the luminosity of the primary component requires an evolutionary model with an upper mass limit extending beyond the substellar boundary .Models that are constrained to the substellar regime are therefore unsuitable for this system, as they do not extend to the higher masses required to describe objects near or above the stellar boundary. By contrast, both the BCAH15 and SM08 models provide broader mass ranges that encompass objects above the Brown Dwarf regime. These models also offer complementary approaches to atmospheric physics: the BCAH15 models assume cloudless atmospheres, whereas the SM08 models account for cloud formation. This distinction enables us to examine the impact of clouds on objects straddling the stellar/substellar regime.

\begin{figure}
\begin{center}
\includegraphics[clip,width=\columnwidth]{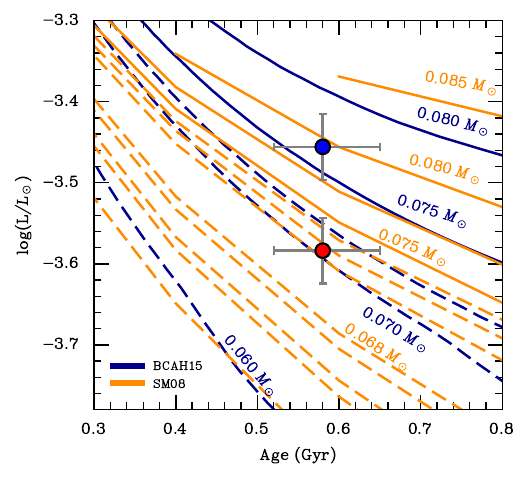}
\caption{Age-luminosity plane with the detected companions J1250+0455A (blue marker) and J1250+0455B (red marker) with iso-mass tracks taken from the SM08 (orange) and BCAH15 (blue) model grids.} 
\label{iso_mass_plot}
\end{center}
\end{figure}

\begin{figure}
\begin{center}
\includegraphics[clip,width=\columnwidth]{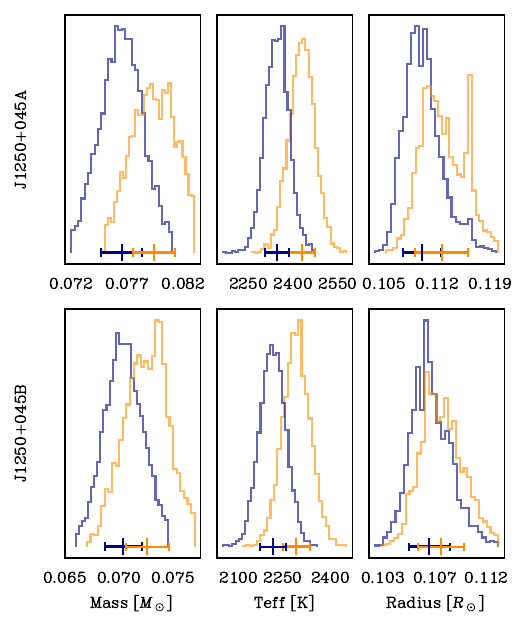}
\caption{Mass, effective temperature and radius constraints for J1250+0455A (top row) and J1250+0455B (bottom row) obtained using SM08 (orange) and BCAH15 (blue). The horizontal bars indicate the median and 1$\sigma$ error for each model in their respective colours for each distribution. } 
\label{mass_teff_radii_plot}
\end{center}
\end{figure}

By leveraging the model-independent age and luminosity, we estimated the mass of the components by comparing their positions in the age-luminosity plane to iso-mass tracks derived from the SM08 and BCAH15 models, as shown in Fig.\ref{iso_mass_plot}. A bootstrapping method was employed to propagate the posterior distributions of age and luminosity to derive meaningful estimates. This was achieved through linear interpolation between isomass tracks using \texttt{scipy.interpolate.griddata} \citep{2020SciPy-NMeth}. By drawing 10,000 random samples from the age-luminosity posterior distributions for each component, we derived the posterior distributions for their masses, as shown in Fig. \ref{mass_teff_radii_plot}. Using the same approach, we obtained posterior distributions for the companions' effective temperatures and radii; the results are presented in the middle and right columns of Fig. \ref{mass_teff_radii_plot}. A summary of these findings is presented in Table \ref{J1250+0455_binary_properties}.
Both components lie near the stellar--substellar boundary, with both models slightly favouring a stellar classification for J1250+0455A and a substellar classification for J1250+0455B. However, these classifications should be interpreted with caution due to the likely underestimated uncertainties regarding the age of the system. The derived $T_\mathrm{eff}$ and radii are consistent with J1250+0455A being the larger and more massive component, as expected. The models agree within 1$\sigma$, indicating general consistency despite some discrepancies in the predicted parameters. This alignment supports the current knowledge that the influence of cloud physics on objects near the stellar--substellar transition may be less significant than for lower-mass L/T transition objects.

\subsubsection{Binding energy and orbital period}
\label{BE and orbital period}
 We calculated the projected angular separation between the components of J1250+0455AB in Sec. \ref{binary parameters} as 170 $\pm$ 15\,mas. Using the parallax for J1250+0455AB provided by Gaia DR3 of $\varpi$ = 13.94 $\pm$ 1.14 mas, we derived the projected physical separation between the components to be 12.2 $\pm$ 1.5\,AU. The projected physical separation isthe closest physical separation consistent with the current on-sky separation, in the absence of any additional orbital information. In this study, to estimate the orbital period of the system using Kepler's third law, we adjusted the projected separation to account for potential variations in the orbital inclination and eccentricity. Following the method outlined in \cite{1992ApJ...396..178F}, the projected separation is multiplied by a factor of 1.26. This adjusted separation along with the masses derived in Sec. \ref{mass_teff_radii section}, was used to calculate the orbital period for J1250+0455AB, using both masses derived from SM08 and BHAC15 given in Table \ref{J1250+0455_properties}. A large orbital period implies that the detection of significant orbital motion of the system components is not possible unless long-term monitoring of the system is conducted.\\  The binding energy of system was calculated using :  \begin{equation} E_b = 1.8 \times 10^5 \times \left(\frac{\text{M}_{\text{host}} \times \text{M}_{\text{companion}}}{\text{Separation}}\right) \end{equation}  where M$_{\text{host}}$ refers to the mass of J1250+0455A, and M$_{\text{companion}}$ represents the mass of J1250+0455B, both in M$_\odot$. The separation is measured in AU, and the constant in the formula yields the binding energy in units of 10$^{41}$\,erg \citep{2024AJ....167..253R}.   A binary system remains stable as long as its separation does not exceed the point at which galactic tidal forces surpass the gravitational binding force of the system \citep{1990AJ....100.1968C}. 
 
\begin{figure*}
\begin{center}
\includegraphics[clip,width=\columnwidth]{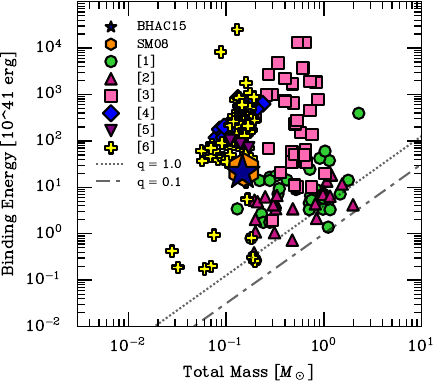}
\includegraphics[clip,width=\columnwidth]{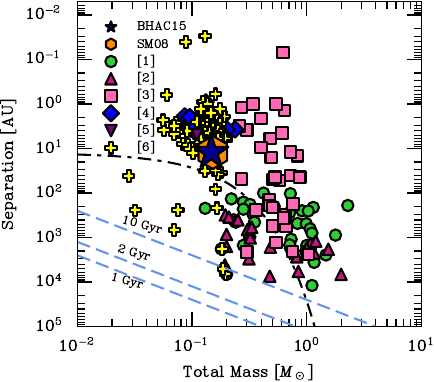}
\caption{\textbf{[Left]}: System binding energy versus total mass. The grey dotted and dash-dotted lines indicate the Jeans length criterion for mass ratios q = 1.0, and q = 0.1, following the approach outlined in \citet{2010AJ....139..176F}. The orange hexagon and blue star markers represent the binding energies of J1250+0455AB calculated using the masses from SM08 (orange hexagon) and BHAC15 (blue star). \textbf{[Right]}: Binary separation versus total mass. The plot includes lifetime isochrones adapted from \citet{1987ApJ...312..367W}, using data from \citet{2007ApJ...660.1492C} and detailed in \citet{2010AJ....139.2566D} for dissipation times of 1, 2, and 10 Gyr (blue dashed lines). The empirical stability limit described by \citet{2001AJ....121..489R} is indicated by a black dash-dotted line. Both plots include various other binary systems represented by distinct markers. The legend provides the following references: [1]: \citet{2010AJ....139..176F}, [2]: \citet{2015ApJ...802...37B}, [3]: \citet{1992ApJ...396..178F}, [4]: \citet{2017ApJS..231...15D}, [5]: \citet{2005ApJ...633..452K}, and [6]: Very Low Mass Binaries Archive, \citet{2007prpl.conf..427B}. }
\label{BE_plot}
\end{center}
\end{figure*}

 To assess the stability of J1250+0455AB and the likelihood of disruption, we examined the empirical relations for binary stability in Fig.\ref{BE_plot}. An empirical binding energy threshold for systems formed by fragmentation was derived in \cite{2009A&A...493.1149Z}, assuming a binary separation of 300 AU. However, this limit was revised by \cite{2010AJ....139..176F}, as numerous systems with separations greater than 300 AU were observed. \cite{2010AJ....139..176F} applied the Jeans length criterion, recalculating the limits for two mass-ratio cases, q=1.0, and q=0.1, both of which are represented in Fig.\ref{BE_plot}. The binding energy of J1250+0455AB lies well above these thresholds, indicating a stable binary configuration. Moreover, J1250+0455AB demonstrates binding energies comparable to those of other low-mass binaries (M$_\odot$$_\mathrm{tot}$ < 0.2M$_\odot$) at similar separations, particularly those discussed in \cite{2010AJ....139..176F},  \cite{2007prpl.conf..427B} and \cite{2005ApJ...633..452K}.      
We also present the systems in \cite{2017ApJS..231...15D} which have a total mass similar to that of J1250+0455AB. While having a similar total mass these systems have masses that are dynamically determined and thus have tighter orbits, smaller separations and hence higher binding energies.
 
 \cite{2001AJ....121..489R} provided a statistical model for the maximum separation that a binary system can maintain over a given age, based on the data available at that time. However, as more widely separated systems have been discovered, this model has proven less applicable. More recent work by \cite{2010AJ....139.2566D} refines this relationship by incorporating the galactic disk mass density from \cite{2007ApJ...660.1492C} and equations from \cite{1987ApJ...312..367W} \citep[explained in detail in][]{2023A&A...670A.102G}. The resulting lifetime isochrones for dissipation times of 1 Gyr, 2 Gyr, and 10 Gyr are shown in Fig. \ref{BE_plot}, alongside the earlier model by \cite{2001AJ....121..489R}. J1250+0455AB falls within the anticipated region of the phase space for systems with comparable mass and separation, aligning well with the observed distributions of similar binaries.
 The estimated age and binding energy of J1250+0455AB suggest that the system is both strongly bound and highly stable, and unlikely to be disrupted or dissolved, indicating that the binary is likely to remain intact and survive for more than 10 Gyr.

\subsubsection{Orbital Acceleration in Gaia DR4}
Discrepancies in measured proper motions in the plane of the sky between different epochs 
(``astrometric acceleration'') have proven to be a powerful tool for measuring precise 
dynamical masses of long-period systems 
\citep{2010A&A...509A.103S, 2011A&A...525A..95S, 2019AJ....158..140B, 2019ApJ...871L...4D, 
2020ApJ...904L..25C, 2021AJ....162..301B}. \textit{Gaia}~DR4, which will be based on 
66~months of observations, is expected to release all epoch and transit astrometry for 
every source in its catalogue. This unprecedented access to high-precision relative 
astrometry opens new avenues for dynamical mass determinations, especially for 
low-mass multiple systems that are otherwise challenging to characterise via radial 
velocities, or direct imaging.

We estimate the astrometric acceleration for J1250+0455AB using the approximate formalism of 
\citep[Eq.~15 of ][]{1999PASP..111..169T}. Adopting a companion mass of 
$M_B \approx 0.071\,M_\odot$, a distance $D \approx 71\,\mathrm{pc}$, and an angular 
separation $\rho \approx 0.17''$, we simplify this by setting $\Psi \approx 1$. The factor 
$\Psi(i,e,\omega,\Omega,\varphi)$ depends on the orbital geometry and eccentricity 
(including the inclination $i$, eccentricity $e$, argument of pericentre $\omega$, 
longitude of the node $\Omega$, and orbital phase $\varphi$). For an order-of-magnitude 
estimate, we take $\Psi=1$ to represent a ``median'' orientation, that is,\ 
avoiding highly face-on or edge-on configurations or extreme eccentricities. This yields 
a characteristic acceleration of 
$\sim 270\,\mu\mathrm{as}\,\mathrm{yr}^{-2}$. Given that \textit{Gaia}~DR4 may detect 
accelerations on the order of tens to a few hundred $\,\mu\mathrm{as}\,\mathrm{yr}^{-2}$ at 
$G\approx20$, it is plausible that next-generation \textit{Gaia} astrometry could measure 
(or at least constrain) the orbital motion of J1250+0455AB, thereby helping to refine its 
dynamical mass estimates.

\begin{table}
\caption{Derived parameters for J1250+0455A and J1250+0455B. The photometric properties include H- and K$_\text{s}$-band magnitudes and bolometric luminosity. Astrometric properties include projected angular and physical separation, and position angle. Astrophysical properties include spectral type (photometry and spectral fitting), gyrochronological age, mass, effective temperature, radius, orbital period, and binding energy derived using SM08 and BCAH15 evolutionary models.}
\label{J1250+0455_binary_properties}
\centering
\begin{tabular}{@{}lccc@{}}
\toprule
Parameter (Unit) & J1250+0455A & J1250+0455B \\ \midrule
\multicolumn{3}{c}{\textbf{Derived Photometry}} \\ \midrule
H (mag) & 15.02 $\pm$ 0.06 & 15.32 $\pm$ 0.06 \\
K$_\text{s}$(mag) & 14.50 $\pm$ 0.05 & 14.76 $\pm$ 0.05 \\
log(L/L$_\odot$)  & -3.45 $\pm$ 0.04 & -3.58 $\pm$ 0.04 \\ \midrule
\multicolumn{3}{c}{\textbf{Derived Astrometry}} \\ \midrule
Sep (mas) & \multicolumn{2}{c}{170 $\pm$ 15} \\
Sep (AU) & \multicolumn{2}{c}{12.2 $\pm$ 1.5} \\
PA ($^\circ$) & \multicolumn{2}{c}{84.8 $\pm$ 0.2} \\ \midrule
\multicolumn{3}{c}{\textbf{Derived Astrophysical Properties}} \\ \midrule
SpT$^{\text{Phot}}$ (-) & M9.5 $\pm$ 0.6 & L0.5 $\pm$ 0.6 \\
SpT$^{\text{SF}}$ (-) & M9 $\pm$ 1.1 & L0 $\pm$ 1.4 \\
Age (Gyr) & \multicolumn{2}{c}{$0.58^{+0.07}_{-0.06}$} \\
M$^{\text{SM08}}$ (M$_\odot$) & 0.079 $\pm$ 0.002 & 0.072 $\pm$ 0.003 \\
M$^{\text{BCAH15}}$ (M$_\odot$) & 0.077 $\pm$ 0.002 & 0.070 $\pm$ 0.003 \\
T$_\text{eff}^{\text{SM08}}$ (K) & 2350 $\pm$ 38 & 2220 $\pm$ 43 \\
T$_\text{eff}^{\text{BCAH15}}$ (K) & 2430 $\pm$ 42 & 2300 $\pm$ 38 \\
R$^{\text{SM08}}$ (R$_\odot$) & 0.113 $\pm$ 0.003 & 0.108 $\pm$ 0.002 \\
R$^{\text{BCAH15}}$ (R$_\odot$) & 0.110 $\pm$ 0.003 & 0.107 $\pm$ 0.002 \\
P$^{\text{SM08}}$ (yrs) & \multicolumn{2}{c}{155 $\pm$ 9} \\
P$^{\text{BCAH15}}$ (yrs) & \multicolumn{2}{c}{157 $\pm$ 8} \\
U$^{\text{SM08}}$ (erg) & \multicolumn{2}{c}{$67.1 \times 10^{41}$} \\
U$^{\text{BCAH15}}$ (erg) & \multicolumn{2}{c}{$62.9 \times 10^{41}$} \\ \bottomrule
\end{tabular}
\end{table}

\section{Conclusions}
\label{conclusion}
In this study, we present a detailed analysis of the newly discovered UCD binary system J1250+0455AB. The system was resolved using high-resolution NIR imaging from the LUCI instrument on the LBT, which revealed two near-equal-mass components. Our analysis of the newly discovered binary is summarised below : 

\begin{itemize} 
\item We calculate a projected physical separation of 12.2 $\pm$ 1.5,AU between the components, with a PA of 84.8 $\pm$ 0.2$^\circ$. A flux ratio of 1.27 $\pm$ 0.01 indicates near-equal masses. 
\item Both photometric estimates and spectral fitting suggest the primary has a spectral type of M9.5/L0.5 and the secondary L0/L1. 
\item TESS data for the primary component of the hierarchical system J1250+04553 reveals a rotational period of 11.48 days, corresponding to an estimated age of $0.56^{+0.07}_{-0.06}$ Gyr using gyrochronological relations of \cite{2024AJ....167..159L}. We adopt this as the age of J1250+0455AB, assuming that the system is coeval.
\item Both components straddle the stellar-substellar boundary, with an average mass of the primary of 0.078 $\pm$ 0.02 M$_\odot$, while the secondary has a mass of 0.071 $\pm$ 0.01 M$_\odot$. 
\item The effective temperature of J1250+0455A is 2390 $\pm$ 71\,K and J1250+0455B is 2255$\pm$ 75\,K. The radii of the components are 0.113 $\pm$ 0.003\,R$_\odot$ and 0.108 $\pm$ 0.002\,R$_\odot$, respectively.

 \item The estimated orbital period of J1250+0455AB is 156 $\pm$ 8 years, making immediate dynamical mass estimates challenging without long-term astrometric monitoring. The orbital motion can be measured with epoch astrometry from $\gaia$ DR4, providing an avenue for dynamical mass determination of the system. \item The system has a binding energy of 65 × 10$^{41}$ erg and is firmly bound and stable. The binary should remain bound and stable beyond 10\,Gyr. 
\end{itemize}

J1250+0455AB adds to the small but growing group of resolved triple systems containing at least one UCD, as reported in \citep{1994A&A...291L..47L, 2005AJ....129..511B, 2020MNRAS.499.5302D, 2024MNRAS.533.3784B}. With a model-independent age, the system serves as a valuable age benchmark, particularly within the stellar--substellar mass regime. Systems hosting UCDs can exhibit notably divergent ages or mass-luminosity relationships. For example, in the 2MASS~J0700+3157AB system the fainter component is more massive, leading to SM08 model ages for the primary and secondary components which cannot be easily reconciled under coeval formation \citep{2019ApJ...871L...4D}. This illustrates how standard mass--luminosity relations may yield inconsistent results, especially for borderline objects that can be mischaracterised if one component is unresolved or if its atmosphere significantly deviates from conventional assumptions. By contrast, hierarchical triple systems can provide a reliable age reference when at least one stellar companion is well characterised, thereby enabling more precise constraints. 

Acquiring high-resolution spectra of J1250+04553 is essential for accurately determining the chemical composition of J1250+0455AB. These spectra would establish a reliable chemical benchmark for the system by leveraging the well-characterised composition of the brighter host star. This is particularly important because UCDs are intrinsically faint, which makes direct compositional measurements challenging. By deriving the chemical properties of the host, we can model the companion’s atmosphere with greater precision, offering potential insights into cloud formation, condensation, and atmospheric dynamics.

The efficacy of our method in identifying unresolved UCD binaries opens up larger sample sizes for follow-up AO observations. This is particularly relevant for known systems such as J1250+0455AB, which already resides in a multiple system. $\gaia$ DR4  is expected to offer many new candidates, benefiting from improved astrometry and inclusion of epoch and transit data, thus enabling precise dynamical mass estimations for the tightest companions.

\section*{Acknowledgements}
This study was supported by the Science and Technology Facilities Council through a PhD studentship to SB (ST/W507490/1) and through research infrastructure support to SB and HRAJ (ST/V000624/1). RLS has been supported by a STSM grant from  COST Action CA18104: MW-Gaia.

We acknowledge the support from the LBT-Italian Coordination Facility for the execution of observations, data distribution and reduction. Observations have benefited from the use of ALTA Center (\href{alta.arcetri.inaf.it}{alta.arcetri.inaf.it}) forecasts performed with the Astro-Meso-Nh model. Initialization data of the ALTA automatic forecast system come from the General Circulation Model (HRES) of the European Centre for Medium Range Weather Forecasts.

We thank Federico Marocco, J. Davy Kirkpatrick, and Rocio Kiman for their valuable discussions and insightful advice, particularly on the topics of rotation periods and age dating of M-dwarfs.

This research has made use of the NASA Exoplanet Archive, which is operated by the California Institute of Technology, under contract with the National Aeronautics and Space Administration under the Exoplanet Exploration Program. 
We also acknowledge the valuable contributions of TOPCAT \citep{2005ASPC..347...29T, 2006ASPC..351..666T}, Scipy \citep{2020SciPy-NMeth}, Astropy \citep{2013A&A...558A..33A}, the VizieR catalogue access tool and the SIMBAD database operated at CDS, Strasbourg, France; National Aeronautics and Space Administration (NASA) Astrophysics Data System (ADS). The $\gaia$ mission has been pivotal for this work provided by the European Space Agency (ESA) \href{https://www.cosmos.esa.int/gaia}{(https://www.cosmos.esa.int/gaia)} processed by the Gaia Data Processing and Analysis Consortium (DPAC, \href{https://www.cosmos.esa.int/web/gaia/dpac/consortium}{(https://www.cosmos.esa.int/web/gaia/dpac/consortium)}. Funding for the DPAC has
been provided by national institutions, in particular, the institutions participating
in the Gaia Multilateral Agreement.

The LBT is an international collaboration among institutions in the United States, Italy and Germany. LBT Corporation partners are: The University of Arizona on behalf of the Arizona university system; Istituto Nazionale di Astrofisica, Italy; LBT Beteiligungsgesellschaft, Germany, representing the Max-Planck Society, the Astrophysical Institute Potsdam, and Heidelberg University; The Ohio State University, and The Research Corporation, on behalf of The University of Notre Dame, University of Minnesota, and University of Virginia,
We acknowledge the support from the LBT-Italian Coordination Facility for the execution of observations, data distribution and reduction
\section*{Data Availability}
The data from the LBT is hosted on the \href{https://archive.lbto.org/}{LBT archive} under proposal ID: IT-2023B-035. The BHAC15 models can be accessed \href{https://perso.ens-lyon.fr/isabelle.baraffe/}{here} and SM08 models \href{https://www.carolinemorley.com/models}{here}.

\bibliographystyle{mnras}
\bibliography{references}
\bsp	
\label{lastpage}
\end{document}